\documentclass{article}
\usepackage{spconf,amsmath,graphicx}
\usepackage{amsmath,bm}
\usepackage{floatrow}
\usepackage{multirow}
\usepackage{multicol}
\usepackage{subfigure}
\usepackage{setspace}
\usepackage{microtype}
\newfloatcommand{capbtabbox}{table}[][\FBwidth]

\title{ENHANCEMENT OF A CNN-BASED DENOISER BASED ON SPATIAL AND SPECTRAL ANALYSIS}
\name{Rui Zhao, Kin-Man Lam, and Daniel P.K. Lun}
\address{Department of Electronic and Information Engineering\\
The Hong Kong Polytechnic University\\
rick10.zhao@connect.polyu.hk, \{enkmlam, enpklun\}@polyu.edu.hk\\
}
\begin{document}

%
\maketitle
\begin{abstract}
Convolutional neural network (CNN)-based image denoising methods have been widely studied recently, because of their high-speed processing capability and good visual quality. However, most of the existing CNN-based denoisers learn the image prior from the spatial domain, and suffer from the problem of spatially variant noise, which limits their performance in real-world image denoising tasks. In this paper, we propose a discrete wavelet denoising CNN (WDnCNN), which restores images corrupted by various noise with a single model. Since most of the content or energy of natural images resides in the low-frequency spectrum, their transformed coefficients in the frequency domain are highly imbalanced. To address this issue, we present a band normalization module (BNM) to normalize the coefficients from different parts of the frequency spectrum. Moreover, we employ a band discriminative training (BDT) criterion to enhance the model regression. We evaluate the proposed WDnCNN, and compare it with other state-of-the-art denoisers. Experimental results show that WDnCNN achieves promising performance in both synthetic and real noise reduction, making it a potential solution to many practical image denoising applications.
\end{abstract}


\section{Introduction}
\label{sec:intro}
Current image capturing technologies may result in additional noises to the acquired images, which greatly affects their perceptual quality and the performance of further high-level computer vision tasks \cite{DIR}. Thus, image denoising is an important step before other operations. Image denoising is an ill-posed task, which aims to estimate the clean image $\mathbf{x}$ from its noisy observation $\mathbf{y}$, which follows the common degradation model $\mathbf{y} = \mathbf{x} + \mathbf{n}$, where $\mathbf{n}$ is assumed to be an additive white Gaussian noise with a standard deviation $\sigma_n$. Current state-of-the-art denoisers, from a Bayesian point of view, attempt to model or learn some image priors. They can be divided into two categories, model-based methods and learning-based methods. The non-local self-similarity (NNS) is one of the most effective priors, such that many methods, based on this prior, achieved remarkable performance, such as BM3D \cite{BM3D} and WNNM \cite{WNNM}. Although the model-based denoisers can deal with the corrupted images with different noise levels, they all suffer from the problems that i) the optimization process in these methods is computationally intensive, and ii) multiple hyper-parameters need to be set manually, making the algorithms highly complicated.
\vspace{-0.33pt}

The learning-based methods, instead of modeling handcrafted priors, learn the underlying deep image prior from the noisy and noise-free image pairs. These methods usually train a denoiser in the spatial domain, such as DnCNN \cite{DnCNN}, FFDNet \cite{FFDNet}, and FTMD \cite{FTMF}. However, spectral analysis is a more effective approach to remove noise. Since noise mainly exhibits as high-frequency components, removing noise in the frequency spectra is much more reliable and efficient. Wavelet transformation is one of the most commonly used spectral analysis tools and its coefficients contain both spatial and spectral information, making it an effective technique for image restoration. Recently, more and more research has been focusing on building the connections between CNNs and wavelet transformation. Fujieda \emph{et al.} \cite{WCNN} proposed a wavelet convolutional neural networks (WCNN), in which multiresolution wavelet transformation was adopted to generate enhanced features, which cannot be learnt by CNNs. Furthermore, Liu \emph{et al.} \cite{MWCNN} replaced the normal pooling layers in the CNNs with the wavelet packet transformations, in order to preserve more texture details and achieve desirable results. 
\vspace{-0.33pt}

However, most of the information in natural images mainly resides in the low-frequency spectrum, resulting in the magnitude of the low-band coefficients being much larger than that of the high-band coefficients. This highly imbalanced property makes the hidden feature maps established principally by the low-band coefficients, and greatly affects the reconstruction of the detailed textures. Therefore, in this paper, we propose a band normalization module (BNM) to normalize the scale of the different-band coefficients. Furthermore, a band discriminative training (BDT) criterion is applied to the proposed wavelet denoising CNN (WDnCNN), for achieving better regression of each sub-band. Details of BNM and BDT will be discussed in Sections 2 and 3, respectively.

\section{PROPOSED DISCRETE WAVELET DENOISING CNN MODEL}
\label{sec:PROPOSED discrete WAVELET DENOISING CNN MODEL}

Linear, orthogonal transforms are of great interests in image denoising, because, by linearity, the additive behaviour of noises can be maintained. Specifically, the common degradation model can be transformed into the spectral domain as $\mathbf{u = w + v}$, where $\mathbf{u}$, $\mathbf{w}$, and $\mathbf{v}$ are the noisy signal, clean signal, and the additive noise in the spectral domain, respectively. By orthogonality, the statistical behaviour of the noise can be maintained as well. Therefore, based on the Parseval relation \cite{Parseval}, we have
\begin{equation}
    \setlength\abovedisplayskip{5pt}
    \setlength\belowdisplayskip{5pt}
    \mathbf{||\hat{w} - w||}^2 = \mathbf{||\hat{x} - x||}^2,
\end{equation}
which implies that denoising can be performed completely in the transformed domain. In this paper, the proposed WDnCNN serves as a mapping function $\mathbf{\hat{w}} = \mathcal{F}(\mathbf{u}, \bm{M}; \Theta)$, where $\bm{M}$, inspired by \cite{FFDNet}, is the noise level map, which is of the same size as $\mathbf{u}$ with all elements equal to $\sigma_n$, and $\Theta$ is the deep image prior. Our proposed network is able to handle noisy images with different noise levels.

\subsection{Network architecture}
\label{subsec: Network Architecture}
\begin{figure*}[ht]
    \centering
    \includegraphics[width=0.90\linewidth]{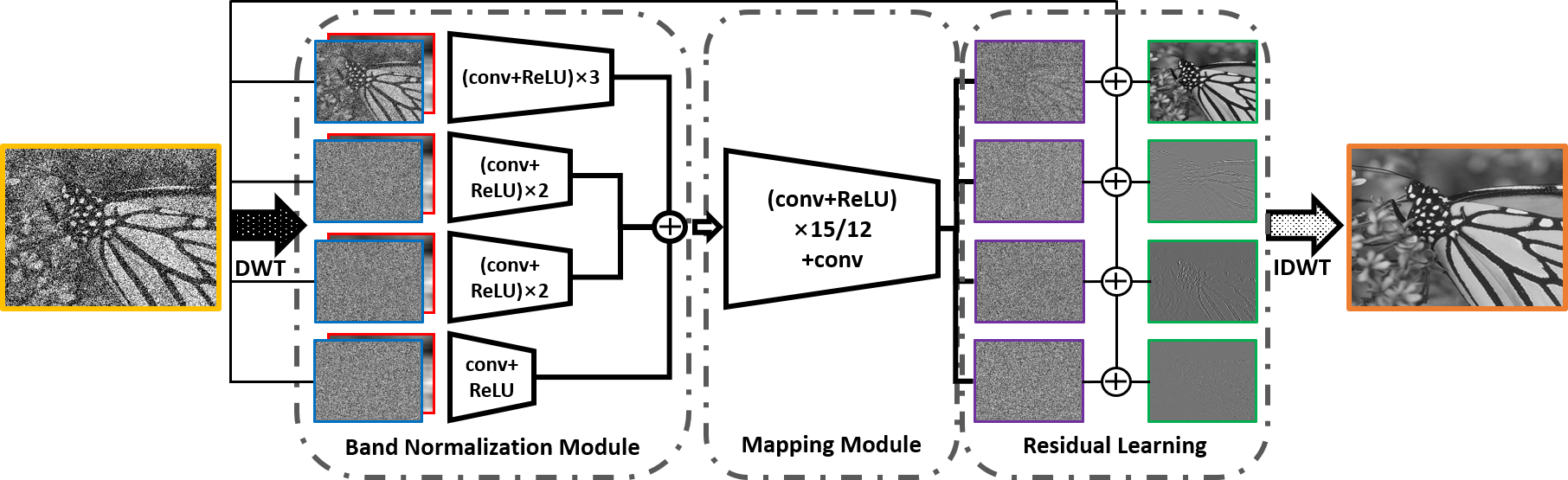}
    \vspace{-10pt}
    \caption{The architecture of the proposed WDnCNN. The dark arrow represents the discrete wavelet transform (DWT), and the light arrow represents the inverse discrete wavelet transform (IDWT).}
    \label{fig:1}
\end{figure*}
The whole network architecture is shown in \textbf{Fig. \ref{fig:1}}. The network takes a noisy observation of size $W \times H \times C$ as input, where $W$, $H$, and $C$ are the width, the height, and the channels, respectively, i.e. $C=1$ for grayscale images and $C=3$ for color images. After the discrete wavelet transform (DWT), it becomes a tensor $\bm{t}$ of size $W_0 \times H_0 \times 4C$, where $W_0 \leq W$ and $H_0 \leq H$. $W_0$ and $H_0$ are determined by the size of the input image and the length of the wavelet filter. Each sub-band tensor $\bm{t}_i$ is then concatenated with a uniform noise level map, whose elements are all equal to $\sigma_n$, and the size of the sub-band tensors becomes $W_0 \times H_0 \times (C+1)$. We send these sub-band tensors to the corresponding branches of the band normalization module for training.

Since WDnCNN is a fully convolutional network without pooling layers, and the output of WDnCNN is of the same size as the input, we set all convolutional filters in WDnCNN to the size of $3 \times 3$ and the padding size to $1$. Different from DnCNN and FFDNet, we do not apply any batch normalization (BatchNorm). The reason for this will be discussed in Sec. 2.3. Considering the scales and the semantic intensity of different filter bands, we set the number of convolutional layers in BNM to $3$, $2$, $2$, and $1$ for the LL, LH, HL, and HH bands, respectively. Considering both effectiveness and efficiency of the network, we set the number of convolutional layers in the mapping module to 16 and 13 for grayscale and color images, respectively. In terms of the number of channels for feature maps, we set 72 for grayscale images and 108 for color images, because the color images contain more information and require more feature maps for description. Inspired by DnCNN, we adopt residual learning to train the network. The residual learning aims to learn the pixel-wise noise for each sub-band. It is applied because the high-frequency bands contain more information about noise, which provides a better capacity for the network to predict the residual.

\subsection{Band normalization plus noise level map}
\label{subsec: Band normalization plus noise level map}

By using handcrafted image priors for denoising, we aim to solve the problem as follow:
\begin{equation}
    \setlength\abovedisplayskip{5pt}
    \setlength\belowdisplayskip{5pt}
    \mathbf{\hat{w}} = \mathop{\arg\min}_{\mathbf{w}} ||\mathbf{u}-\mathbf{w}||^2 + \lambda\Theta(\mathbf{w}),
\end{equation}
where $\mathbf{w}$ and $\mathbf{u}$ can be the clean and noisy image pair in any domain. $\Theta(\mathbf{w})$ is the regularization term, which relates to the image prior, and $\lambda$ is the weight that controls the balance between noise removal and detailed texture protection \cite{FFDNet}. It has been proven that the weight-control parameter $\lambda$ is highly related to the noise level $\sigma_n$. Therefore, it justifies that concatenating a noise level map in FFDNet to guide the pixel-wise noise reduction is reasonable. However, for denoising in the spectral domain, each of the frequency spectra should be handled unequally, by setting $\lambda$ to different values. Thus, the above problem can be changed to
\begin{equation}
    \setlength\abovedisplayskip{5pt}
    \setlength\belowdisplayskip{5pt}
    \mathbf{\hat{w}}_k = \mathop{\arg\min}_{\mathbf{w}_k} ||\mathbf{u}_k-\mathbf{w}_k||^2 + \lambda_k\Theta(\mathbf{w}_k),
\end{equation}
where $\mathbf{w}_k$ and $\mathbf{u}_k$ are the clean and noisy pair of the $k$-th frequency component, and $\lambda_k$ is the weight of the $k$-th frequency component. To discriminatively guide each band for regression, we propose an unequal-depth structure in the band normalization module to generate the different $\lambda_k$. 

In the band normalization module, unequal-depth branches are applied to the different band coefficients. Therefore, using more convolutional layers for the LL band and less convolutional layers for the HH band can benefit scale unification. Semantically, we need more convolutional layers for the lower frequency components to learn their noise distribution. By introducing the band normalization module with an unequal-depth structure, we achieve to use i) a single denoising network for different noise levels, and ii) unequal weights in different sub-bands for better texture preservation.

\subsection{Non BatchNorm of CNN}
In DnCNN \cite{DnCNN}, internal covariate shift (ICS) \cite{BN} was addressed as the main problem for not obtaining sufficient regression. Therefore, batch normalization is applied to reduce the impact of ICS. Other networks based on DnCNN (e.g. FFDNet and MWCNN) also inherit this structure. However, in \cite{HBN}, it was found that the relation between BatchNorm and ICS reduction is tenuous, and BatchNorm might even increase ICS in training. The main contribution made by BatchNorm is that it makes the loss gradients more reliable. Therefore, considering the efficiency of our denoising model, we do not add any batch normalization layers, but propose a training criterion, namely band discriminative training (BDT), to enhance the regression of WDnCNN.

\section{EXPERIMENTS}
\label{sec:EXPERIMENTS}

\subsection{Dataset generation}
To train our model, we collect 4,744 images from the Waterloo Exploration Database \cite{Waterloo}, 400 images from the Berkeley segmentation dataset \cite{BSD}, and 600 images from the validation set of ImageNet \cite{ImageNet}, to generate the whole training set for grayscale image denoising. As for color image denoising, we replace 400 images from the Berkeley segmentation dataset with another 400 images from the validation set of ImageNet. For each epoch, we randomly select 1,000 noise-free images from the training set and then randomly crop $128\times 2,000$ patches, with patch size of $50\times 50$, from them. To obtain pairs of clean and noisy patches, additive white Gaussian noise (AWGN) is added to the selected patches. This is because the noise in a local region of a real noisy image can be assumed as AWGN. Moreover, we densely sample the noise level $\sigma_n$ in a range of $[0, 75]$ for each synthetic noisy patch. The mini-batch size is set to 128 for all epochs, and the rotation-flip operations are employed for data augmentation.

We apply the ADAM \cite{ADAM} algorithm to the optimization process for minimizing the following loss function
\begin{equation}
    \setlength\abovedisplayskip{5pt}
    \setlength\belowdisplayskip{5pt}
    \mathcal{L}(\Theta) = \frac{1}{2KN}\sum_{k=1}^{K}\sum_{i=1}^{N}\mu_k||\mathcal{F}_k(\mathbf{u}_{ik},\bm{M}_{ik};\Theta)-\mathbf{w}_{ik}||^2,
\end{equation}
where $K$ is the number of filter bands, and $\mu_k$ represents the weight for the $k$-th band. For the wavelet transformation, we simply choose the discrete Meyer wavelet (dmey) filter, because of its rapid decay and compact support in the frequency domain. Moreover, it does not cause any aliasing problems or distortions in the reconstruction process.

\subsection{Band discriminative training}
\begin{figure}[ht]
    \centering
    \includegraphics[width=0.88\linewidth]{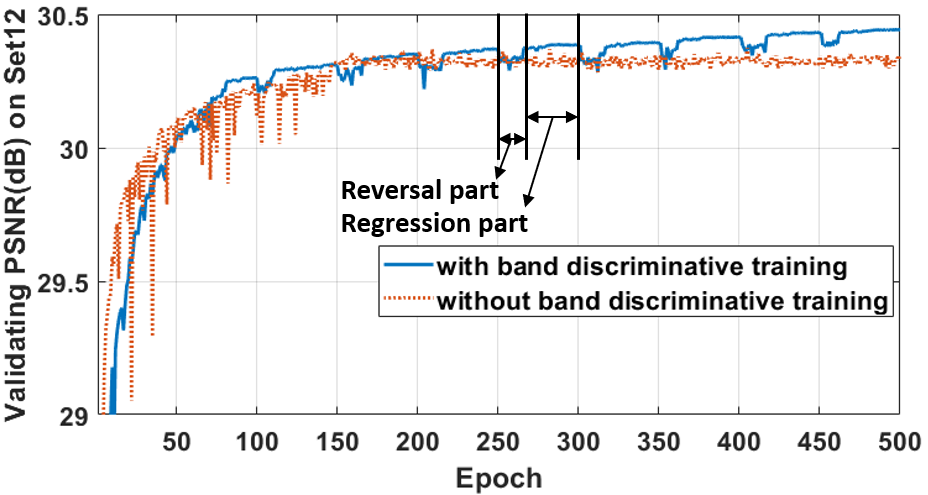}
    \vspace{-15pt}
    \caption{BDT vs Non-BDT of WDnCNN.}
    \label{fig:2}
\end{figure}

To achieve a better regression for each sub-band, we initialize the convolutional filters with Kai\_ming\_normal \cite{KaiMing} in Pytorch. Due to the unequal importance of the different bands, conventional training procedures cannot lead to sufficient regression. To tackle this problem, we first train the initialized model with $\mu_k = 1.5$, $2.5$, $2.5$, and $5$ for the LL, LH, HL, and HH bands, respectively. Because the wavelet coefficients of the clean image are unnormalized, we use these inversely proportional weights for keeping the target coefficients in balance. The learning rate is kept at $10^{-4}$ until the model fully converges. Then, we modify $\mu_k$ as shown in \textbf{Table \ref{tab:1}}, for each 50 epochs, and fine-tune our model with a learning rate decreasing from $10^{-4}$ to $10^{-7}$. Each fine-tuning process can be divided into a reversal part and a regression part. A comparison of BDT and Non-BDT is shown in \textbf{Fig. \ref{fig:2}}. It is clear that BDT provides a more stable training process and a better regression performance.
\vspace{-10pt}
\begin{table}[ht]
    \small
    \centering
    \begin{tabular}{c|c|c|c}
          Epoch & (LL, LH, HL, HH) & Epoch & (LL, LH, HL, HH)\\
         \hline
         1-50      & (2.0, 2.5, 2.5, 4.5) & 50-100  & (3.5, 2.5, 2.5, 3.0) \\
         100-150   & (4.5, 2.5, 2.5, 2.0) & 150-200 & (5.5, 1.5, 1.5, 1.0) \\
         200-250   & (6.0, 2.0, 2.0, 1.5) & 250-300 & (6.5, 2.5, 2.5, 2.0) \\
         300-350   & (7.0, 3.0, 3.0, 2.5) & 350-400 & (7.5, 3.5, 3.5, 3.0) \\
         400-450   & (8.0, 4.0, 4.0, 3.5) & 450-500 & (8.5, 4.5, 4.5, 4.0)
    \end{tabular}
    \caption{The band discriminative training weights.}
    \vspace{-15pt}
    \label{tab:1}
\end{table}

\vspace{-15pt}
\begin{table*}[t]
    \small
    \centering
    \begin{tabular}{|c|c|c|c|c|c|c|c|c|c|c|c|c|c|}
        \hline
        Images & C.man & House & Pepp. & Star. & Mona. & Airp. & Parr. & Lena & Barb. & Boat & Man & Couple & \emph{Ave.}\\
        \hline
        \multicolumn{14}{|c|}{$\sigma_n=25$}\cr\cline{1-14}
        \hline
        BM3D & 29.45 & 32.85 & 30.16 & 28.56 & 29.25 & 28.42 & 28.93 & 32.07 & 30.71 & 29.90 & 29.61 & 29.71 & 29.969 \\
        \hline
        WNNM & 29.64 & 33.22 & 30.42 & 29.03 & 29.84 & 28.69 & 29.15 & 32.24 & \textbf{31.24} & 30.03 & 29.76 & 29.82 & 30.257 \\
        \hline
        FFDNet & 30.06 & 33.27 & 30.79 & \textbf{29.33} & 30.14 & \textbf{29.05} & \textbf{29.43} & 32.59 & 29.98 & 30.23 & \textbf{30.10} & 30.18 & 30.429 \\
        \hline
        Ours & \textbf{30.12} & \textbf{33.33} & \textbf{30.90} & 29.20 & \textbf{30.18} & 29.01 & 29.41 & \textbf{32.65} & 29.99 & \textbf{30.31} & 30.08 & \textbf{30.23} & \textbf{30.451} \\
        \hline
        \multicolumn{14}{|c|}{$\sigma_n=50$}\cr\cline{1-14}
        \hline
        BM3D & 26.13 & 29.69 & 26.68 & 25.05 & 25.82 & 25.10 & 25.90 & 29.05 & 27.22 & 26.78 & 26.81 & 26.46 & 26.722 \\
        \hline
        WNNM & 26.45 & 30.33 & 26.95 & 25.44 & 26.32 & 25.42 & 26.14 & 29.25 & \textbf{27.79} & 26.97 & 26.94 & 26.64 & 27.052 \\
        \hline
        FFDNet & 27.03 & 30.43 & 27.43 & \textbf{25.77} & \textbf{26.88} & \textbf{25.90} & 26.58 & 29.68 & 26.48 & 27.32 & \textbf{27.30} & 27.07 & 27.322 \\
        \hline
        Ours & \textbf{27.34} & \textbf{30.55} & \textbf{27.65} & 25.61 & 28.86 & 25.86 & \textbf{26.60} & \textbf{29.75} & 26.51 & \textbf{27.38} & 27.27 & \textbf{27.14} & \textbf{27.376} \\
        \hline
        \multicolumn{14}{|c|}{$\sigma_n=75$}\cr\cline{1-14}
        \hline
        BM3D & 24.32 & 27.51 & 24.73 & 23.27 & 23.91 & 23.48 & 24.18 & 27.25 & 25.12 & 25.12 & 25.32 & 24.70 & 24.909 \\
        \hline
        WNNM & 24.60 & 28.24 & 24.96 & 23.49 & 24.31 & 23.74 & 24.43 & 27.54 & \textbf{25.81} & 25.29 & 25.42 & 24.86 & 25.224 \\
        \hline
        FFDNet & 25.29 & 28.43 & 25.39 & \textbf{23.82} & \textbf{24.99} & 24.18 & 24.94 & 27.97 & 24.24 & 25.64 & 25.74 & 25.29 & 25.493 \\
        \hline
        Ours & \textbf{25.75} & \textbf{28.60} & \textbf{25.73} & 23.59 & 24.91 & \textbf{24.22} & \textbf{24.97} & \textbf{28.00} & 24.40 & \textbf{25.72} & \textbf{25.75} & \textbf{25.39} & \textbf{25.584} \\
        \hline
    \end{tabular}
    \caption{The PSNR(dB) results on Set12 by the different methods. The best results are highlighted in \textbf{bold}.}
    \vspace{-15pt}
    \label{tab:3}
\end{table*}

\begin{table}[ht]
    \small
    \centering
    \begin{tabular}{c|c|c|c|c}
        Datasets & $\sigma_n$ & CBM3D & FFDNet & Ours\\
        \hline
        \multirow{3}{*}{CBSD68}  & 25 & 30.71 & 31.21 & \textbf{31.24}\\
                                 & 50 & 27.38 & 27.96 & \textbf{28.07}\\
                                 & 75 & 25.74 & 26.24 & \textbf{26.39}\\
        \hline
        \multirow{3}{*}{Kodak24} & 25 & 31.68 & 32.13 & \textbf{32.23}\\
                                 & 50 & 28.46 & 28.98 & \textbf{29.15}\\
                                 & 75 & 26.82 & 27.27 & \textbf{27.47}\\
        \hline
        \multirow{3}{*}{McMaster}& 25 & 31.66 & 32.35 & \textbf{32.44}\\
                                 & 50 & 28.51 & 29.18 & \textbf{29.41}\\
                                 & 75 & 26.79 & 27.33 & \textbf{27.62}\\
    \end{tabular}
    \caption{The average PSNR(dB) results on color images by the different methods. The best results are highlighted in \textbf{bold}.}
    \vspace{-15pt}
    \label{tab:2}
\end{table}

\subsection{Synthetic and real-world noise removal}

We evaluate our denoising model on the Set12, CBSD68 \cite{BSD}, McMaster \cite{MM}, and Kodak24 \cite{Kodak24} datasets for synthetic noise reduction. The tested noisy images are obtained from the above datasets corrupted by AWGN with noise level $\sigma_n$ equal to 25, 50, and 75. We compared our proposed method with other state-of-the-art methods, such as BM3D \cite{BM3D} (or CBM3D for color images), WNNM \cite{WNNM}, and FFDNet \cite{FFDNet}. \textbf{Table \ref{tab:3}} and \textbf{Table \ref{tab:2}} tabulate the results, in terms of PSNR(dB), of the different methods on grayscale and color images, respectively. The state-of-the-art denoisers selected for our experiments can handle different noise levels with a single model.

It can be observed from the results that WDnCNN outperforms the benchmark BM3D by a large margin on all testing sets. Compared to WNNM, WDnCNN can obtain a PSNR gain of about 0.3dB for denoising grayscale images. Although WDnCNN only achieves comparable results with FFDNet when the noise level is low($\sigma_n \leq 25$), it shows great ability in distinguishing texture information from noise, contributing to a relatively larger PSNR improvement on \emph{C.man}, \emph{Peppers}, and \emph{House}. Furthermore, WDnCNN can regress the noise distribution more discriminatively on different sub-bands, making it more effective when noise is stronger and able to achieve a PSNR gain of about 0.2dB, with $\sigma_n = 75$, compared to FFDNet. Considering the fact that we used a long wavelet filter, the performance of WDnCNN can be further improved by using Sym8 for a larger receptive field.

\begin{figure}[ht]
    \subfigure[real-world noisy image]{
    \includegraphics[width=0.45\linewidth]{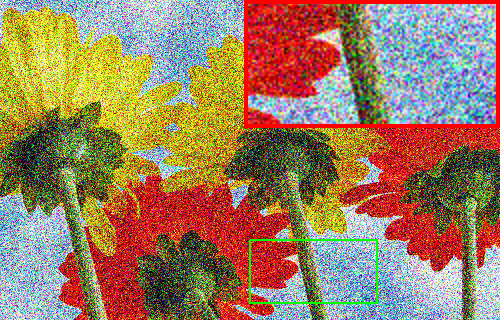}
    }
    \subfigure[Denoised by CBM3D]{
    \includegraphics[width=0.45\linewidth]{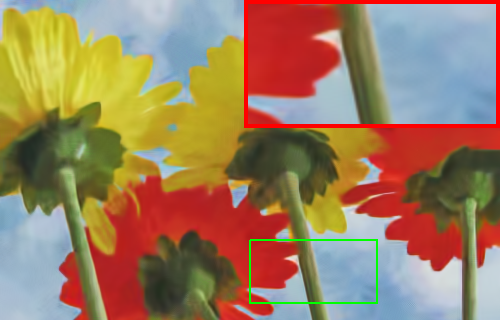}
    }
    \subfigure[Denoised by FFDNet]{
    \includegraphics[width=0.45\linewidth]{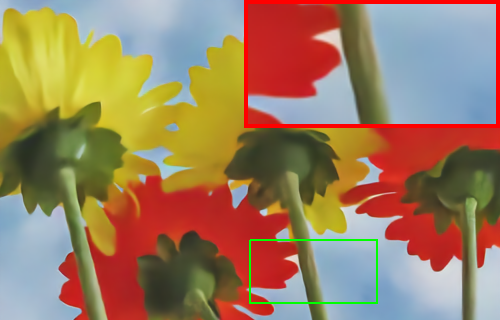}
    }
    \subfigure[Denoised by WDnCNN]{
    \includegraphics[width=0.45\linewidth]{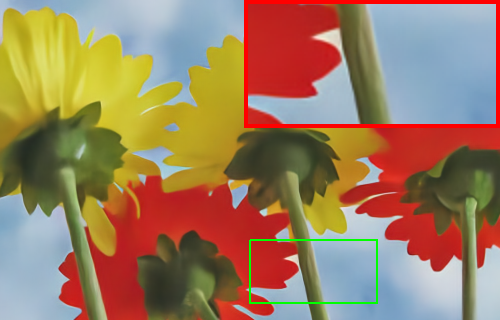}
    }
    \vspace{-20pt}
    \caption{Visual qualities of a restored real-world noisy image by the different methods.}
    \label{fig:3}
\end{figure}

To further evaluate WDnCNN, we tested it on some real-world noisy images. Since the ground-truth clean images are unavailable, visual comparison is adopted for assessment. One example is shown in \textbf{Fig. \ref{fig:3}}. It can be seen that WDnCNN produces a better, detailed texture preservation, as the grain and wrinkles on the flower stalk are more vivid. In summary, WDnCNN establishes state-of-the-art visual quality for the restored images.
\vspace{-5pt}
\vspace{-20pt}

\section{CONCLUSION}
\label{sec:CONCLUSION}
In this paper, we proposed a new denoising model, named WDnCNN. By introducing a band normalization module and a band discriminative training criterion, WDnCNN can achieve a better regression of noise in the spatial-spectral domain. Extensive experiments demonstrated that WDnCNN achieves promising performance on both synthetic and real-world noise reduction with a single model, making it a potential solution to many practical image denoising problems.


\bibliographystyle{IEEEbib.bst}
\bibliography{bibliography.bib}
\end{document}